\documentclass[aps,prd,twocolumn,showkeys,nofootinbib]{revtex4}
\pdfoutput=1
\usepackage{epsfig}
\usepackage{amsmath}
\usepackage{graphicx}
\usepackage{multirow}
\newcommand{\beq}{\begin{equation}}
\newcommand{\eeq}{\end{equation}}
\newcommand{\bqa}{\begin{eqnarray}}
\newcommand{\eqa}{\end{eqnarray}}
\newcommand{\nnb}{\nonumber\\}

\newcommand{\bold}{\textbf}

\def\bfsigma{\mbox{\boldmath $\sigma$}}

\begin{document}

\title{Form factors for semileptonic $B_{c}$  decays into $\eta^{(')}$ and Glueball}
%\title{Gluonium contributions and the form factors of $B_{c}$ transitions into $\eta^{(')}$ }
\author{Ruilin Zhu} \email{rlzhu@njnu.edu.cn}
\author{Yan Ma }
\author{Xin-Ling Han}
\author{Zhen-Jun Xiao}\email{xiaozhenjun@njnu.edu.cn}
\affiliation{
 Department of Physics and Institute of Theoretical Physics,
Nanjing Normal University, Nanjing, Jiangsu 210023, China\\
 }
%\date{\today}
\begin{abstract}
We calculated the  form factors of $B_{c}$ transitions into $\eta^{(')}$ meson and pseudoscalar  Glueball, where the $B_{c}$  meson is a bound state of two different heavy
 flavors and is treated as a nonrelativistic state,  while the mesons
  $\eta^{(')}$ and Glueball
are treated as  light-cone objects since their masses  are smaller enough compared to the transition momentum scale.
The mechanism of two gluon scattering into $\eta^{(')}$  dominated
 the form factors of $B_{c}$  decays into $\eta^{(')}$.
 We considered the $\eta-\eta'-$Glueball mixing effect, and then obtained their
  influences on the form factors. The form factors of $B_{c}$ transitions
  into $\eta^{(')}$ and pseudoscalar Glueball in the maximum momentum  recoil point were obtained
  as follows: $f^\eta_{0,+}(q^2=0)=1.38^{+0.00}_{-0.02}\times 10^{-3}$,
  $f^{\eta'}_{0,+}(q^2=0)=0.89^{+0.11}_{-0.10}\times 10^{-2}$ and  $f^{G}_{0,+}(q^2=0)=0.44^{+0.13}_{-0.05}\times 10^{-2}$.
  Also phenomenological discussions for semileptonic $B_{c}\to
  \eta^{(')}+\ell+\bar{\nu}_\ell$,  $B_{c}\to
 G(0^{-+})+\ell+\bar{\nu}_\ell$ and $D_{s}\to \eta+\ell+\bar{\nu}_\ell$ decays are given.

\end{abstract}
%\pacs{12.38.Bx,  13.25.Gv, 14.40.Pq}

\maketitle

\section{Introduction}

Hadron-hadron colliders currently provide the unique platform to investigate the production and decay properties of the $B_c$ meson as the bound state of two different heavy flavors. In pace with the running of the CERN Large Hadron Collider (LHC) with  the luminosity of about ${\cal L}\sim 10^{33} cm^{-2}s^{-1}$, one can expect around $10^9$ $B_c$ events per year~\cite{Chang:2003cr}. When a tremendous amount of $B_c$ events are reconstructed, one can systematically and precisely test the golden decay channels of the $B_c$ meson or hunt for its rare decays~\cite{Qiao:2012hp}.

The $B^-_c$ meson has two different heavy flavors and its decay modes can be classified into three categories:
(i) the anti-charm quark decays with $\bar{c}\to \bar{d}, \bar{s}$;
(ii) the bottom quark decays with $b\to u, c$;
and (iii) the weak annihilation where both the bottom and anti-charm decays.
These three categories of decay modes contribute to the total decay width of the $B^-_c$ meson are around 70, 20, and 10 percent, respectively~\cite{Kar:2013fna}.
There are currently  a lot of  theoretical and experimental works on the singly heavy quark decays of the $B_c$ meson,
some of which can be found
in Refs.~\cite{Chang:1992pt,Ebert:2003cn,Wang:2008xt,Dubnicka:2017job,Zhu:2017lwi,Zhu:2017lqu,Qiao:2012hp,He:2016xvd}.
And the studies of the rare weak annihilation decays of  the $B_c$ meson are few, some of which can be
 found in Refs.~\cite{Gang:2006nc,Xiao:2011zz,Liu:2017cwl,Wang:2015bka,Chen:2015csa,Zhu:2018epc}.

In this paper, we will investigate the decay properties of the $B_c$ meson
into the light pseudoscalar mesons $\eta$, $\eta'$ and Glueball. The light pseudoscalar
mesons with quark contents are organized into two representations: singlet and octet according to flavor
SU(3) symmetry. Due to Isospin symmetry, the form factors of the $B_c$ meson  into
Isospin triplet $\pi^0$ is trivial, where the contributions from the quark contents
$u\bar{u}$ and $d\bar{d}$ in $\pi^0$ will be cancelled out. Thus the form factors of the $B_c$ meson
 into the light meson $\pi^0$ will only depend on a small Isospin symmetry breaking effect,
 which is very similar to the case where the cross sections of $e^+e^-\to J/\psi\eta(\eta') $
 are around $pb$ while there is no signal for $e^+e^-\to J/\psi\pi^0 $~\cite{Ablikim:2016ymr,Qiao:2014pfa}.

In the flavor SU(3) symmetry,  the light mesons with quark contents  form the flavor-octet and the flavor-singlet.
The masses of these light mesons become identical and trivial in the limit of zero quark mass.
 For different masses for the light $u$, $d$, and $s$ quarks and the flavor symmetry breaking,
 the light mesons in the flavor-octet and the flavor-singlet will gain their masses.
  On the other hand, the axial $U(1)$ anomaly will lead to a large mass difference between the $\eta$
   and $\eta'$, which can not be ignored~\cite{Rosenzweig:1981cu,Feldmann:1999uf,Shore:2007yn,Kawarabayashi:1980dp,Mathieu:2009sg}.
    Besides,  the flavor singlet and octet contents, even then the gluonium state will be mixed with each
    other with the identical $J^{PC}$ and form the physical $\eta^{(')}$
     states~\cite{Rosenzweig:1981cu,Feldmann:1999uf,Shore:2007yn,Kawarabayashi:1980dp,Mathieu:2009sg,Zhu:2015qoa}.  The $\eta$ meson is viewed as the mixing state between flavor singlet and octet contents, and the gluonium content is usually suppressed. However the conventional singlet-octet basis is not enough to explain the content of $\eta'$. For example, the gluonium contribution reached few percents in   $B\to\eta'$ decays. Thus  the $\eta '$ is viewed as
the mixing state among $q\bar{q}$, $s\bar{s}$, and $gg$.

The $B_c$ meson is treated as a nonrelativistic bound state, where the heavy quark relative velocity is small in the rest frame of the meson. The nonrelativistic QCD (NRQCD) effective theory is employed to deal with the decays of
the $B_c$ meson. Considering the  mass of the light meson $P$  is less than the $B_c$ meson,
i.e. $m^2_{P}<<m_{B_{c}}^2$, a large momentum is transferred in the $B_{c}$ transitions into the light meson $P$. The light meson $P$ can be treated as
a light cone object in the rest frame of the $B_c$ meson. In the maximum momentum recoil point with
$q^2=0$,  the form factors of the $B_{c}$ transitions into the light meson $P$ can be factored as
the hadron long-distance matrix elements and the corresponding perturbative short distance coefficients.

We will discuss the properties of the form factors of the $B_{c}$ transitions into the light pseudoscalar mesons $\eta$, $\eta'$ and Glueball.  We will employ the form factors formulae into the related semileptonic  decays, naming $B_{c}\to \eta^{(')}+\ell+\bar{\nu}_\ell$ and $B_{c}\to
 G(0^{-+})+\ell+\bar{\nu}_\ell$.

The paper is organized as the following. In Sec.~\ref{II}, we will introduce the NRQCD approach, the
$\eta-\eta'-$Glueball mixing effect, the light cone distribution amplitudes and the scattering mechanism of two gluons into light mesons.
In Sec.~\ref{III}, we will calculate the form factors of the $B_{c}$ meson into $\eta^{(')}$ and Glueball.
 Especially, we will determine the quark-antiquark pair and gluonium contributions to the form factors and discuss their properties.
In Sec.~\ref{IV}, we will study the semileptonic decays of the $B_{c}$ meson into $\eta^{(')}$ and Glueball. And we will tentatively analyze the processes $D_{s}\to \eta+\ell+\bar{\nu}_\ell$.
We summarize and conclude in the end.

\section{Factorization formulae\label{II}}

\subsection{NRQCD effective theory }
The heavy quark  relative velocity is a small quantity  inside  the heavy quarkonium and then the heavy quark pair is nonrelativistic in the rest frame of heavy
quarkonium. The quark
relative velocity  squared is estimated as $v^2\approx 0.3$ for $J/\psi$ and $v^2\approx 0.1$
for $\Upsilon$~\cite{Bodwin:1994jh}. The $B_c$ meson is usually treated as a nonrelativitic state
and the quark reduced velocity squared is estimated in the region $0.1 <v^2<0.3$. The calculations of the productions and decays
of the heavy quarkonium and the $B_c$ meson with a large momentum transmitted  usually  refer to
the NRQCD effective theory
established by Bodwin, Braaten, and Lepage~\cite{Bodwin:1994jh}.

In the NRQCD effective theory, the Lagrangian is written as~\cite{Bodwin:1994jh}
\begin{eqnarray}
%----------------------
{\mathcal L}_{\rm NRQCD} &=&
\psi^\dagger \left( i D_t + {{\bf D}^2 \over 2m} \right) \psi
+ {c_F \over 2 m} \psi^\dagger \bfsigma \cdot g_s {\bf B} \psi
%----------------------
\nonumber\\
%----------------------
&+& \psi^\dagger {{\bf D}^4 \over 8m^3} \psi+{c_D\over 8 m^2} \psi^\dagger ({\bf D}\cdot g_s {\bf E}- g_s {\bf E}\cdot {\bf D})\psi
%----------------------
\nonumber\\
%----------------------
&+&{i c_S\over 8 m^2} \psi^\dagger \bfsigma \cdot ({\bf D}\times g_s {\bf E}- g_s {\bf E}\times {\bf D})\psi
%----------------------
\nonumber\\
%----------------------
&+& \left(\psi \rightarrow i \sigma ^2 \chi^*, A_\mu \rightarrow - A_\mu^T\right) +
{\mathcal L}_{\rm light} \,,
%----------------------
\label{NRQCD:Lag}
%----------------------
\end{eqnarray}
%----------------------
where
$\psi$ and $\chi$ represent the two-component Pauli spinor field that annihilates a heavy quark and creates a heavy antiquark with quark mass $m$, respectively. $\bfsigma $ is the Pauli matrix.  The electric and magnetic color components of the gluon field strength tensor $G^{\mu\nu}$ are denoted as $E^i=G^{0i}$ and $B^i=\frac{1}{2}\epsilon^{ijk}G^{jk}$, respectively. The space and  time components of the gauge-covariant derivative $D^\mu$ are denoted as $\bf D$ and $D_t$, respectively.   ${\mathcal L}_{\rm light}$ denotes the Lagrangian for the light quarks and gluons. The short-distance coefficients $c_D$, $c_F$, and $c_S$ can be perturbatively calculated according to the matching procedure between QCD and NRQCD calculations.

Within the framework of NRQCD, the heavy quarkonium inclusive annihilation decay width is factorized as~\cite{Bodwin:1994jh}
\begin{eqnarray}
\Gamma(H)=\sum_n\frac{2\mathrm{Im}f_n(\mu_\Lambda)}{m^{d_n-4}}\langle H|{\cal O}_n(\mu_\Lambda)|H\rangle\,,
\end{eqnarray}
where$\langle H|{\cal O}_n(\mu_\Lambda)|H\rangle$  are  NRQCD decay long-distance matrix elements (LDMEs) , which involve nonperturbative information
 and obey the power counting rules, which are ordered by the  relative velocity between the heavy quark and antiquark
inside the heavy quarkonium $H$.

The leading order  NRQCD decay operators for the decay of $S$-wave heavy quarkonium   are
\begin{eqnarray}
\mathcal{O}(^{1}S_{0}^{[1]})&=&\psi^{\dagger}\chi\chi^{\dagger}\psi,\\
\mathcal{O}(^{3}S_{0}^{[1]})&=&\psi^{\dagger}\bfsigma\chi\cdot\chi^{\dagger}\bfsigma\psi.
\end{eqnarray}
These operators are also valid for the $B_c$ family with two different heavy flavors.

For a certain process, the matching coefficients multiplying decay LDMEs are determined through perturbative
matching between QCD and NRQCD at the amplitude level.  The covariant projection method is another equivalent but more convenient approach to extract the short-distance coefficients of the NRQCD LDMEs. The corresponding projection operators are defined as
\begin{widetext}
\begin{eqnarray}
\Pi_{S=0,1}(k) &=&  -i\sum_{\lambda_1,\lambda_2} u_1(p_1,\lambda_1)\bar{v}_2(p_2,\lambda_2)\langle\frac{1}{2}\lambda_1\frac{1}{2}\lambda_2|S S_z\rangle\otimes \{\frac{\bold{1}_c}{\sqrt{N_c}}, \sqrt{2}\bold{T}^a\}\nonumber\\
&=&\frac{i}{4\sqrt{2 E_1 E_2}\omega}(\alpha \,p\!\!\!\slash_{H}-k\!\!\!\slash+m_1)\frac{p\!\!\!\slash_{H}+E_1+E_2}{E_1+E_2}\Gamma_S
(\beta\,p\!\!\!\slash_{H}+k\!\!\!\slash-m_2)\otimes \{\frac{\bold{1}_c}{\sqrt{N_c}}, \sqrt{2}\bold{T}^a\}\,,\label{projection}
\end{eqnarray}
\end{widetext}
where $\omega=\sqrt{E_1+m_1}\sqrt{E_2+m_2}$ with $E_1=\sqrt{m_1^2-k^2}=\sqrt{m_1^2+\mathbf{k}^2}$ and $E_2=\sqrt{m_2^2-k^2}=\sqrt{m_2^2+\mathbf{k}^2}$. The parameter $\alpha$ and $\beta$ satisfy the relations as $\alpha=E_1/(E_1+E_2)$,  and $\beta=1-\alpha$. We have the spin $S=0$ and $\Gamma_{S=0}=\gamma^5$ for the spin-singlet combination. For the spin-triplet combination, we have the spin $S=1$ and $\Gamma_{S=1}=\varepsilon\!\!\!\slash_{H}=\varepsilon_\mu(p_H) \gamma^\mu$.  $\{\frac{\bold{1}_c}{\sqrt{N_c}}, \sqrt{2}\bold{T}^a\}$ denote
the color-singlet and color-octet projection in the $SU(3)$ color space. For the decays of $B_c^-$, $p_H$ is the $B_c^-$ meson momentum; $p_1=\alpha p_H-k$ is the bottom quark momentum with the mass $m_1=m_b$; $p_2=\beta p_H+k$ is the anti-charm quark momentum with the mass $m_2=m_c$; $k$ is half of the relative momentum between the anti-charm and bottom quarks with  $k^2=-\mathbf{k}^2$.

The heavy quarkonium state is not limited to heavy quark pairs in a color singlet configuration according to NRQCD.  The  heavy quark pairs in a color singlet configuration is only the leading order of Fock state of the quarkonium. Other Fock states sometimes paly an important role
in the inclusive production of heavy quarkonium. In the form factors of the $B_{c}$ meson into  the light mesons, the dominant contribution is from the
color singlet configuration.

\subsection{$\eta -\eta '-$Glueball mixing schemes}

 The $\eta -\eta '-$Glueball mixing effects are discussed in lots of literatures. The popular mixing schemes which are widely employed in these literatures are  the quark-flavor  bases~\cite{Ball:1995zv,Feldmann:1998vh,Escribano:2007cd,Escribano:2008rq,Mathieu:2009sg,Ke:2011fj} and the flavor singlet-octet bases~\cite{Bramon:1997mf,Escribano:2005qq,Cheng:2008ss,Li:2007ky,Liu:2012ib,Xiao:2014uza,Harland-Lang:2017mse}. In the  quark-flavor scheme, the basic quark components are $\eta_q=q\bar{q}=(u\bar{u}+d\bar{d})/\sqrt{2}$ and
$\eta_s=s\bar{s}$.  While the basic flavor components become $\eta_1=(u\bar{u}+d\bar{d}+s\bar{s})/\sqrt{3}$ and
$\eta_8=(u\bar{u}+d\bar{d}-2s\bar{s})/\sqrt{6}$ for the flavor singlet-octet scheme. In addition, the gluonium state $\eta_g=gg$ is introduced  when it has the identical quantum numbers as the two light quark states. For a $\eta -\eta '-$Glueball mixing with identical spin-parity $J^{PC}=0^{-+}$, one has
\begin{eqnarray}
\left (
\begin{array}{c}
|\eta \rangle \\
|\eta^{\prime} \rangle \\
|G \rangle \\
\end{array}
\right ) = U( \phi,\phi_G)\left (
\begin{array}{c}
|\eta_q \rangle \\
|\eta_{s} \rangle \\
|\eta_{g} \rangle \\
\end{array}
\right )  \; , \label{mixqf}
\end{eqnarray}
with the matrix~\footnote{Here we do not consider the mixing between the $\eta$ and Glueball, which
is consistent with the experimental constraints of the production and decays of $\eta$~\cite{Ball:1995zv,Feldmann:1998vh,Escribano:2007cd,Escribano:2008rq,Mathieu:2009sg,
Ke:2011fj,Bramon:1997mf,Escribano:2005qq,Cheng:2008ss,Li:2007ky,Liu:2012ib,Xiao:2014uza,Harland-Lang:2017mse}. If  considering the mixing between the $\eta$ and Glueball, one has to introduce an additional mixing angle, and the mixing matrix will have not zero element.  }
\begin{eqnarray}
U( \phi,\phi_G)= \left (
\begin{array}{ccc}
\cos \phi & -\sin \phi&0 \\
\sin \phi\cos\phi_G & \cos \phi\cos \phi_G & \sin \phi_G  \\
-\sin \phi\sin\phi_G & -\cos \phi\sin \phi_G & \cos \phi_G  \\
\end{array}
\right ) \; . \label{mixm}
\end{eqnarray}
The QCD states $\eta_i$ with $i=q,s,g$ form the physical mass eigenstates $\eta$, $\eta '$ and Glueball.  One candidate for the physical $J^{PC}=0^{-+}$ Glueball state is the $\eta(1405)$, where the corresponding analysis was performed in Ref.~\cite{Cheng:2008ss}. Here we assume that the physic $\eta$ state does not
mix with Glueball, under which two mixing angles $\phi$ and $\phi_G$ are sufficient.

Another equivalent mixing approach for the flavor singlet-octet scheme can be easily obtained by the replacements of $\eta_q\to\eta_8$, $\eta_s\to\eta_1$,  $\phi \to\theta$ and
 $\phi_G \to\phi'_G$. The small angle $\phi_G=\phi'_G$ is adopted  for simplication in the literatures~\cite{Escribano:2008rq,Ke:2011fj,Cheng:2008ss}.  When $\sin\phi_G\to0$, the mixing only occurs between the two quark states.

Considering that the flavor singlet and octet states can be decomposed into the quark flavor states, one has
\begin{eqnarray}
\left (
\begin{array}{c}
|\eta_8 \rangle \\
|\eta_1 \rangle \\
|\eta_g \rangle \\
\end{array}
\right ) = \left (
\begin{array}{ccc}
\cos \theta_i & -\sin  \theta_i&0 \\
\sin  \theta_i & \cos  \theta_i & 0 \\
0& 0 & 1  \\
\end{array}
\right )\left (
\begin{array}{c}
|\eta_q \rangle \\
|\eta_{s} \rangle \\
|\eta_{g} \rangle \\
\end{array}
\right )  \; ,
\end{eqnarray}
where
\begin{eqnarray}
&&\cos \theta_i=\sqrt{\frac{1}{3}} ,\;\;\;  \sin \theta_i=\sqrt{\frac{2}{3}}.
\end{eqnarray}
The parameter $\theta_i=\mathrm{arctan}\sqrt{2}\simeq 54.74^0$ is always named as the ideal mixing angle.
From the observations,  vector or tensor mesons mixing angles where the axial vector anomaly plays no role are always
close to this ideal angle.

The relations between two mixing schemes can be easily obtained as
\begin{eqnarray}
\cos \theta&=&\frac{\sqrt{2}\sin\phi+\cos\phi}{\sqrt{3}} ,\\
\sin \theta&=&\frac{\sin\phi-\sqrt{2}\cos\phi}{\sqrt{3}} .
\end{eqnarray}
Equivalently, one can get the mixing angle for the flavor singlet-octet scheme as
 $\theta=\phi-\mathrm{arctan}\sqrt{2}$.

According to the quantum field theory definition, the decay constants of the  mesons are defined as
\begin{eqnarray}
\langle0|\bar{q'}\gamma^\mu\gamma_5 q'|\eta^{(')}(p)\rangle=if^{q'}_{\eta^{(')}}p^\mu,\quad\quad  (q'=q,s),
\end{eqnarray}
where the decay constants of the  mesons are also related to the decay constants of the quark components as
\begin{eqnarray}
&&f^{q}_{\eta} = f_q\cos\phi ,\;\;\; f^{s}_{\eta} = -f_s\sin\phi,\\
&&f^{q}_{\eta '} = f_q \sin\phi\cos\phi_G ,\;\;\; f^{s}_{\eta '} = f_s\cos\phi\cos\phi_G,
\end{eqnarray}
where the relations will turn to the traditional form in Ref.~\cite{Feldmann:1998vh} when $\phi_G\to0$.

 If one defines the meson decay constants through the flavor SU(3) octet and singlet axial vector current as
\begin{eqnarray}
\langle0|J^{\mu,i}_5|\eta^{(')}(p)\rangle=if^{i}_{\eta^{(')}}p^\mu,\quad\quad  (i=8,1),
\end{eqnarray}
the relations of the decay constants of the  mesons  become
\begin{eqnarray}
&&f^{8}_{\eta} = f_8\cos\theta_8 ,\;\;\; f^{1}_{\eta} = -f_1\sin\theta_1,\\
&&f^{8}_{\eta '} = f_8 \sin\theta_8 ,\;\;\; f^{1}_{\eta '} = f_1\cos\theta_1.
\end{eqnarray}
The flavor singlet and octet decay constants can be obtained by the quark flavor decay constants~\cite{Feldmann:1998vh}
\begin{eqnarray}
&&f_8= \sqrt{\frac{f_q^2+2f_s^2}{3}},\;\;\theta_8 = \phi-\mathrm{arctan}(\sqrt{2}f_s/f_q) ,\;\\
&& f_1= \sqrt{\frac{2f_q^2+f_s^2}{3}},\;\;\theta_1 = \phi-\mathrm{arctan}(\sqrt{2}f_q/f_s).
\end{eqnarray}
From the above equations, one would have $\theta_8=\theta_1=\theta=\phi-\mathrm{arctan}\sqrt{2}$ in the strict
flavor SU(3) symmetry where $f_q=f_s$.

There are several experimental methods to extract the values of the mixing angle and the decay constants, which have been discussed in the literatures~\cite{Ball:1995zv,Feldmann:1998vh,Bramon:1997mf,Escribano:2007cd,Escribano:2008rq,Ambrosino:2006gk,Michael:2013gka,Chang:2012gnb,Aaij:2014jna}.

In  Ref.~\cite{Feldmann:1998vh}, the decays of $\eta'\to \rho \gamma$ and $\rho \to \eta \gamma$ are investigated, where the ratio of their
decay widths is given as
\begin{eqnarray}
\frac{\Gamma(\eta'\to \rho \gamma)}{\Gamma(\rho \to \eta \gamma)}=
3\tan^2\phi \cos^2 \phi_G \left(\frac{m_{\eta'} (1-\frac{m^2_{\rho}}{m^2_{\eta'}} )}{m_{\rho} (1-\frac{m^2_{\eta}}{m^2_{\rho}})}\right)^3.
\end{eqnarray}
In the $\eta'\to \rho \gamma$ decays, the contributions from $\eta_s$ and $\eta_g$ components are suppressed.
Inputting the latest PDG results:$Br(\eta'\to \rho \gamma)=(28.9\pm0.5)\%$, $Br(\rho \to \eta \gamma)=(3.00\pm0.21)\times 10^{-4}$, $\Gamma_\rho=149.1\pm0.8$MeV, and $\Gamma_{\eta'}=0.196\pm0.009$MeV~\cite{PDG2018}, the mixing angles  are
extracted as $\tan\phi\cos\phi_G=0.827^{+0.39}_{-0.34}$.

In Ref.~\cite{Escribano:2007cd}, a global analysis of radiative $V\to P\gamma$ and
$P \to V\gamma$ decays was performed to determine the gluon content of the $\eta^{(')}$
mesons. Allowing for gluonium in the $\eta'$, the mixing angles were found to be $\phi=41.4^0\pm1.3^0$
 and $\sin^2\phi_G=0.04\pm0.09$, naming $\tan\phi\cos\phi_G=0.864^{+0.059}_{-0.078}$.

In Ref.~\cite{Ambrosino:2006gk}, the KLOE collaboration measured the mixing angles by looking for the radiative decays $\phi\to \eta^{(')}\gamma$. Ignoring the gluonium contribution, the mixing angle is fitted into $\phi=41.4^0\pm0.3^0_{stat}\pm0.7^0_{syst}\pm0.6^0_{th}$.
Allowing for gluonium in the $\eta'$, they yielded $\phi=39.7^0\pm0.7^0$ and $\sin^2\phi_G=0.14\pm0.04$,  naming $\tan\phi\cos\phi_G=0.770^{+0.037}_{-0.036}$.

The LHCb collaboration recently have fitted the mixing angles as $\phi=(43.5^{+1.4}_{-2.8})^0$ and $\phi_G=(0\pm24.6)^0$ by a study of $B$ or $B_s^0$ meson decays into $J/\psi \eta$ and $J/\psi \eta'$ at proton-proton collisions~\cite{Aaij:2014jna}. Compared to the small angle $\phi_G$ in
Refs.~\cite{Escribano:2007cd,Aaij:2014jna}, a larger mixing angle  $\phi_G$ is obtained in Refs.~\cite{Escribano:2008rq,Ambrosino:2006gk,Kou:2001pm}.

The quark flavor decay constants can be obtained by their two-photon decays. One has~\cite{Feldmann:1998vh}
\begin{eqnarray}
f_q&=&
\frac{5\alpha}{12\sqrt{2}\pi^{3/2}}[\sqrt{\Gamma(\eta\to\gamma\gamma)/m_{\eta}^3}\cos\phi\nonumber\\
&&+
\sqrt{\Gamma(\eta'\to\gamma\gamma)/m_{\eta'}^3}\frac{\sin\phi}{\cos\phi_G}]^{-1},\\
f_s&=&
\frac{\alpha}{12\pi^{3/2}}[-\sqrt{\Gamma(\eta\to\gamma\gamma)/m_{\eta}^3}\sin\phi\nonumber\\
&&+
\sqrt{\Gamma(\eta'\to\gamma\gamma)/m_{\eta'}^3}\frac{\cos\phi}{\cos\phi_G}]^{-1}.
\end{eqnarray}
The decay constant $f_s$ is not well determined in this way and has a large error. To extract  the values of $f_s$, one can use~\cite{Feldmann:1998vh}
 \begin{eqnarray}
\frac{f_s}{f_q}&=&\frac{\sqrt{2}(m^2_{\eta}\cos^2\phi+m^2_{\eta'}\sin^2\phi-m^2_\pi)}{(m^2_{\eta'}-m^2_{\eta})\sin2\phi}.
\end{eqnarray}
We input the latest PDG results:$Br(\eta\to \gamma \gamma)=(38.41\pm0.20)\%$, $Br(\eta'\to  \gamma\gamma)=(2.22\pm0.08)\%$, $\Gamma_{\eta}=1.31\pm0.05$keV, and $\Gamma_{\eta'}=0.196\pm0.009$MeV~\cite{PDG2018}. Imposing the mixing angles  $\phi=41.4^0\pm1.3^0$ and $\sin^2\phi_G=0.04\pm0.09$ in Ref.~\cite{Escribano:2007cd}, the decay constants  are
extracted as $f_q=(1.05\pm0.02)f_\pi$ and $f_s=(1.34\pm0.03)f_\pi$ with $f_\pi=130.4$MeV.
When imposing the mixing angles $\phi=39.7^0\pm0.7^0$ and $\sin^2\phi_G=0.14\pm0.04$ in Ref.~\cite{Ambrosino:2006gk},
 the decay constants become $f_q=(1.03\pm0.02)f_\pi$ and $f_s=(1.28\pm0.02)f_\pi$. We will input these values in the following calculations.
\\

\subsection{Light cone distribution amplitudes }

 The light cone distribution amplitudes (LCDA) of  $\eta_q$ and $\eta_s$ components in
$\eta$ have the form~\cite{Feldmann:1999uf}
\begin{equation}\label{gegen}
   \Phi^{(q,s)}_\eta (x,\mu)=6x\bar x[1+\sum_{n=2,4,\ldots} a_{n}(\mu)\;
   C_{n}^{3/2}(x-\bar x)]\;,
\end{equation}
where  $x$ and $\bar x = 1 - x$ are the momentum fractions of the light quark and antiquark
inside $\eta_{q,s}$, respectively.  $C_{n}^{3/2}$ and the following $C_{n}^{5/2}$ are the Gegenbauer polynomials. $a_n(\mu)$ is obtained  by the scale evolution at leading-order logarithmic accuracy
\begin{equation}
   a_n(\mu)=\left(\frac{\alpha_s(\mu^2)}{\alpha_s(\mu_0^2)}\right)^{
   \frac{\gamma_n}{\beta_0}} a_n(\mu_0)\;,\label{gegen011}
\end{equation}
where $\beta_0= 11 C_A/3 -2n_f/3$ with flavor number $n_f$ and  the anomalous dimension $\gamma_n$ reads as
\begin{equation}\label{gegen}
   \gamma_n=4C_F
   (\psi(n+2)+\gamma_E-\frac{3}{4}-\frac{1}{2(n+1)(n+2)})\,,
\end{equation}
with the digamma function $\psi(n)$.

The evolution of the LCDA of the quark contents will mix with the gluonium state for  $\eta^\prime$.
In Ref.~\cite{Ohrndorf:1981uz}, the evolution equation for the LCDA of the mixing $q\bar{q}$ and gg state has been calculated.
The corresponding light cone distribution amplitudes are~\cite{Muta:1999tc,Ali:2000ci,Kroll:2002nt,Kroll:2013iwa}
\begin{widetext}
\begin{eqnarray}
 && \Phi^{(q,s)}(x, \mu) = 6 x \bar x \left \{ 1 +
\sum_{n=2,4,\ldots}\left [  a^{(q,s)}_n(\mu) +\rho^g_n  a^{(g)}_n(\mu)  \right ]  C_n^{3/2}(x-\bar x)  \right \}
\;,
 \\
 && \Phi^{(g)}  (x, \mu) = x \bar x
\sum_{n=2,4,\ldots}\left [ \rho^{q,s}_n a^{(q,s)}_n(\mu) +
a^{(g)}_n(\mu) \right ]C_{n-1}^{5/2}(x-\bar x)  \;, \label{eq:gg}
\end{eqnarray}
\end{widetext}
\label{gegen01}
where $a^{(q,s;g)}_n(\mu)$ can be obtained  by the scale evolution at leading-order logarithmic accuracy
\begin{eqnarray}
   a^{(q,s)}_n(\mu)&=&a^{(q,s)}_n(\mu_0) \left ( \frac{\alpha_s (\mu^2)}{\alpha_s
(\mu_0^2)} \right )^{-\gamma^n_+}\;,\label{gegen02}\\
 a^{(g)}_n(\mu)&=&a^{(g)}_n(\mu_0)
\left ( \frac{\alpha_s (\mu^2)}{\alpha_s (\mu_0^2)} \right
)^{-\gamma^n_-}\;.\label{gegen03}
\end{eqnarray}
where the parameters $\gamma_{\pm}^n$ and $\rho_n$ are
\begin{eqnarray}
    \gamma_{\pm}^n &=&\frac{1}{2}[ \gamma_{qq}^n+ \gamma_{gg}^n\pm\sqrt{ (\gamma_{qq}^n- \gamma_{gg}^n)^2+4 \gamma_{gq}^n \gamma_{qg}^n}],
\end{eqnarray}
 with
 \begin{eqnarray}
    \gamma_{qq}^n &=&-\frac{\gamma^n}{\beta_0},\\
    \gamma_{gq}^n &=&\frac{n_f}{\beta_0}\frac{2}{(n+1)(n+2)},
    \end{eqnarray}
\begin{eqnarray}
    \gamma_{qg}^n &=&\frac{C_F}{\beta_0}\frac{n(n+3)}{(n+1)(n+2)},\\
     \gamma_{gg}^n &=&\frac{4C_A}{\beta_0}[\frac{2}{(n+1)(n+2)}-\sum_{j=2}^{n+1}\frac{1}{j}-\frac{1}{12}-\frac{n_f}{6C_A}],
     \nonumber\\
\end{eqnarray}
and
\begin{eqnarray}
    \rho_n^g &=&-\frac{1}{6}\frac{Q_n}{1-P_n}, \quad\quad \rho_n^q=6\frac{P_n}{Q_n},\\
    P_n&=&\frac{\gamma_{+}^n-\gamma_{qq}^n}{\gamma_{+}^n-\gamma_{-}^n},\quad \quad Q_n=\frac{\gamma_{qg}^n}{\gamma_{+}^n-\gamma_{-}^n}.
\end{eqnarray}

\subsection{The scattering mechanism of two gluons into light mesons }

For the $B_c$ meson decays into $\eta^{(')}$, the mechanism of two gluons  scattering  into $\eta^{(')}$
will play an important role.
Two gluons  scattering  mechanism is  blind to quark charges and light quark flavors, so
 the amplitude is identical to $q\bar{q}$ and $s\bar{s}$ except the mixing factor and the decay constant.

\begin{figure}[th]
\begin{center}
\includegraphics[width=0.42\textwidth]{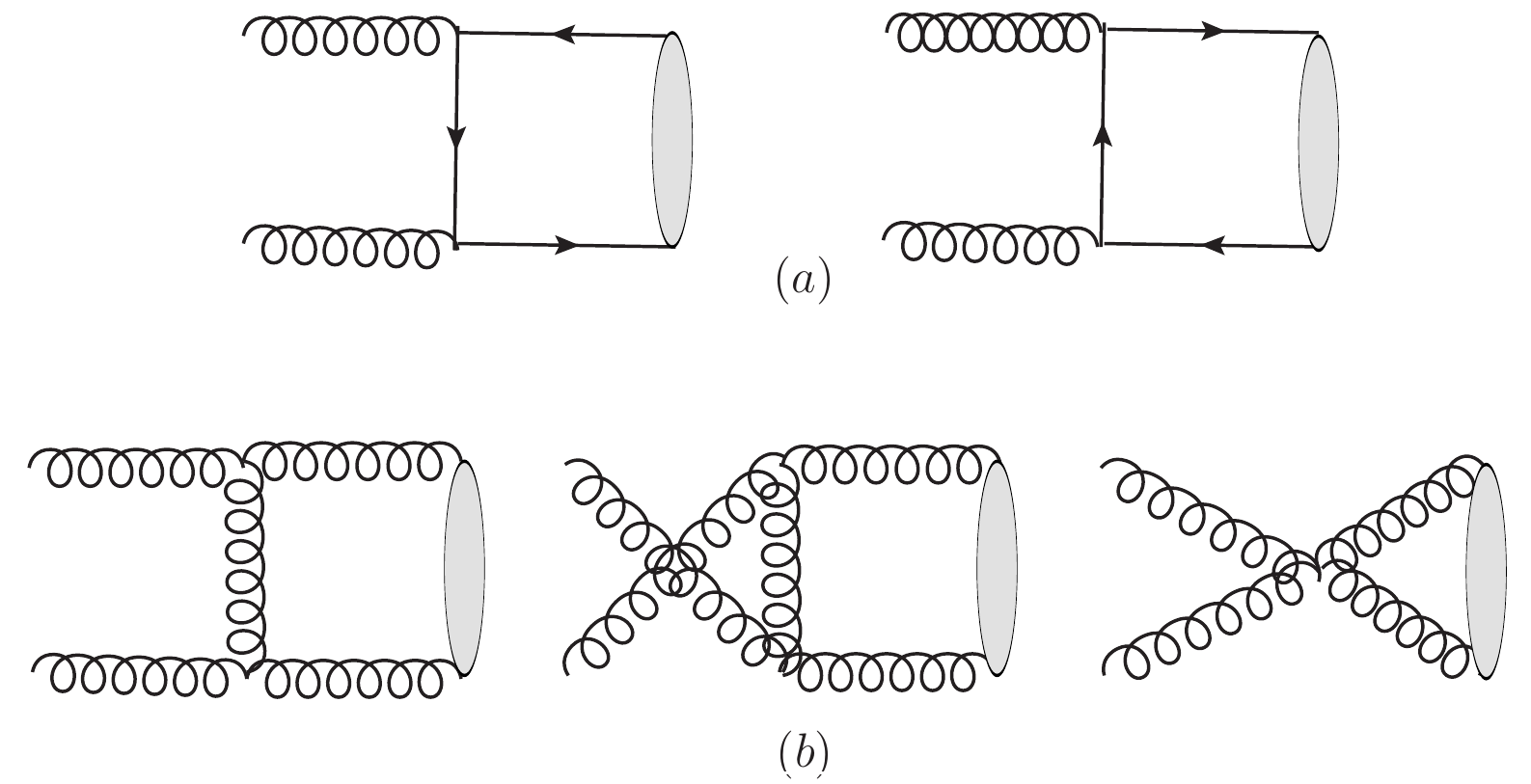}
\end{center}
    \vskip -0.7cm \caption{Feynman diagrams for two gluons scattering into $\eta_q$, $\eta_s$ and $\eta_g$.  }\label{Fig-formfactors}
\end{figure}

The amplitudes of two gluons scattering into quarks and gluonium contents  in lowest-order perturbation theory
can be obtained by calculating the corresponding
 Feynman diagrams which are plotted in Fig.~\ref{Fig-formfactors}.

The amplitude of two gluons scattering to each quark content is
\begin{equation}
{\cal M}^{(q,s)} = -i \, F^{(q,s)}_{ g^* g^*}
(q_1^2, q_2^2) \, \delta_{a b} \,
\varepsilon^{\mu \nu \rho \sigma} \, \varepsilon^{a}_{1\mu}
\varepsilon^{b}_{2\nu} q_{1\rho} q_{2\sigma}, \label{eq:FFQ-1}
\end{equation}
where the momenta of two initial virtual gluons are denoted as $q_1$ and $q_2$,
respectively. The
polarization vectors  of two initial gluons are denoted as $\varepsilon_1(q_1)$ and $\varepsilon_2(q_2)$,
respectively. The form factor $F^{(q,s)}_{ g^* g^*}$  of two gluons transitions to the quark-antiquark content
  can be written as~\cite{Muta:1999tc,Ali:2000ci}
\begin{eqnarray}
&&F^{(q,s)}_{ g^* g^*} (q_1^2, q_2^2) =  \frac{2 \pi \alpha_s (\mu^2)}{ N_c}
\int_0^1 dx \Phi^{(q,s)} (x,
\mu) \,\nonumber
\\&&\times  \left[ \frac{1} {x q_1^2 + \bar x q_2^2 - x \bar x
m_{P}^2 + i \epsilon} + (x \leftrightarrow \bar x)
\right],\label{eq:QFF-2}\nnb
\end{eqnarray}
where $m_P$ is the meson mass.

The amplitude of two gluons scattering into the gluon content can be written as
\begin{equation}
{\cal M}^{(g)} = - i \, F^{(g)}_{g^* g^*} \,
\delta_{a b} \, \varepsilon^{\mu \nu \rho \sigma} \,
\varepsilon^{a}_{1\mu}
\varepsilon^{b}_{2\nu}  q_{1\rho} q_{2\sigma} \,,
\label{eq:FFG-def}
\end{equation}
where the form factors  $F^{(g)}_{ g^* g^*}$  of two gluons scattering to the gluonium content
  can be written as
~\cite{Muta:1999tc,Ali:2000ci}
\begin{eqnarray}
 &&F^{(g)}_{ g^* g^*} (q_1^2, q_2^2)
=\frac{2 \pi \alpha_s (\mu^2)}{Q^2} \,  \int_0^1 dx \,
\Phi^{(g)} (x, \mu)
\nonumber \\
&&  ~~~~\times\left [ \frac{x q_1^2 + \bar x q_2^2 - (1 + x \bar
x) m_{P}^2}
     {\bar x q_1^2 + x q_2^2 - x \bar x m_{P}^2 + i \epsilon}
- (x \leftrightarrow \bar x) \right ], \label{eq:GFF-result1}
\end{eqnarray}
where the typical scale $Q^2$ is introduced to preserve the  dimensionless for
the transition form factors. The choice of $Q^2$ has some freedom, and $Q^2$ is adopted to
$|q_i^2|$ or$|q_1^2+q_2^2|$ in Ref.~\cite{Muta:1999tc,Ali:2000ci}. In this paper,  $Q^2$ is adopted as $m_{P}^2$, and the LCDA of gluonium content
is adopted as the form in Eq.~(\ref{eq:gg}).

For $\eta$, we assume it does not mix with Glueball,  which
is consistent with the experimental constraints~\cite{Ball:1995zv,Feldmann:1998vh,Escribano:2007cd,Escribano:2008rq,Mathieu:2009sg,
Ke:2011fj,Bramon:1997mf,Escribano:2005qq,Cheng:2008ss,Li:2007ky,Liu:2012ib,Xiao:2014uza,Harland-Lang:2017mse}.  The  amplitude of two gluons to $\eta$  is written as
\begin{eqnarray}
{\cal M}^{(q,s)}_\eta
&=&C_\eta {\cal M}^{(q,s)}, \label{eq:etaamp1}
\end{eqnarray}
and the corresponding form factor of two gluons scattering to $\eta$ is
 \begin{eqnarray}
F^{(q,s)}_{\eta g^* g^*} (q_1^2, q_2^2)
&=&C_\eta F^{(q,s)}_{ g^* g^*} (q_1^2, q_2^2). \label{eq:etaFF1}
\end{eqnarray}
 According to the mixing scheme in Eq.~(\ref{mixqf}),
 we have $C_\eta=\sqrt 2 \, f_q\cos\phi -f_s\sin\phi$, which can be viewed as the effective decay constant of $\eta$. In this case, the light meson mass in the formulae becomes $m_P=m_\eta$.

For $\eta'$, the mixing between the quark and  gluon contents should be considered.
The  amplitude of two gluons to $\eta'$  is written as
\begin{eqnarray}
{\cal M}^{(q,s;g)}_{\eta'}
&=&{\cal M}^{(q,s)}_{\eta'}+{\cal M}^{(g)}_{\eta'}
\nonumber\\&=&C^{(q,s)}_{\eta'} {\cal M}^{(q,s)}+C^{(g)}_{\eta'} {\cal M}^{(g)}, \label{eq:etapamp1}
\end{eqnarray}
and the corresponding form factors of two gluons to $\eta'$ are
 \begin{eqnarray}
F^{(q,s)}_{\eta' g^* g^*} (q_1^2, q_2^2)
&=&C^{(q,s)}_{\eta'} F^{(q,s)}_{ g^* g^*} (q_1^2, q_2^2),\\
F^{(g)}_{\eta' g^* g^*} (q_1^2, q_2^2)
&=&C^{(g)}_{\eta'} F^{(g)}_{ g^* g^*} (q_1^2, q_2^2). \label{eq:etapFF1}
\end{eqnarray}
In this case, the light meson mass in the formulae becomes $m_P=m_{\eta'}$.
 According to the mixing scheme in Eq.~(\ref{mixqf}),
 we have  $C^{(q,s)}_{\eta '}=\sqrt 2 \, f_q\sin\phi\cos\phi_G +f_s\cos\phi\cos\phi_G$,  which can be viewed as the effective decay constant of $\eta'$.
  Following the two gluon scattering mechanism proposed in Refs.~\cite{Muta:1999tc,Ali:2000ci}, we do not
  need to introduce the decay constant of $\eta_g$ and we parameterize
 $C^{(g)}_{\eta '}=\sin\phi_G C^{(q,s)}_{\eta '}$.

For Glueball,  the mixing between the gluon and  quark contents should be also considered. The
 amplitude of two gluons to Glueball is written as
\begin{eqnarray}
{\cal M}^{(q,s;g)}_G&=&{\cal M}^{(q,s)}_G+{\cal M}^{(g)}_G\nonumber\\
&=&C^{(q,s)}_{G} {\cal M}^{(q,s)}+C^{(g)}_{G} {\cal M}^{(g)}, \label{eq:Gamp1}
\end{eqnarray}
and the corresponding form factors of two gluons to Glueball are
 \begin{eqnarray}
F^{(q,s)}_{G g^* g^*} (q_1^2, q_2^2)
&=&C^{(q,s)}_{G} F^{(q,s)}_{ g^* g^*} (q_1^2, q_2^2),\\
F^{(g)}_{G g^* g^*} (q_1^2, q_2^2)
&=&C^{(g)}_{G} F^{(g)}_{ g^* g^*} (q_1^2, q_2^2). \label{eq:GFF1}
\end{eqnarray}

In this case, the light meson mass in the formulae becomes $m_P=m_{G}$
 where the candidate of $0^{-+}$ Glueball is $\eta(1405)$ with $m_{\eta(1405)}=1408.8\pm1.8$MeV~\cite{Cheng:2008ss}.
 According to the mixing scheme in Eq.~(\ref{mixqf}) and the two gluon scattering mechanism in Refs.~\cite{Muta:1999tc,Ali:2000ci},
 we have  $C^{(q,s)}_{G}=-\sqrt 2 \, f_q\cos\phi\sin\phi_G -f_s\sin\phi\sin\phi_G$ and  $C^{(g)}_{G}=\cos\phi_G C^{(q,s)}_{G}$.

\begin{figure}[th]
\begin{center}
\includegraphics[width=0.40\textwidth]{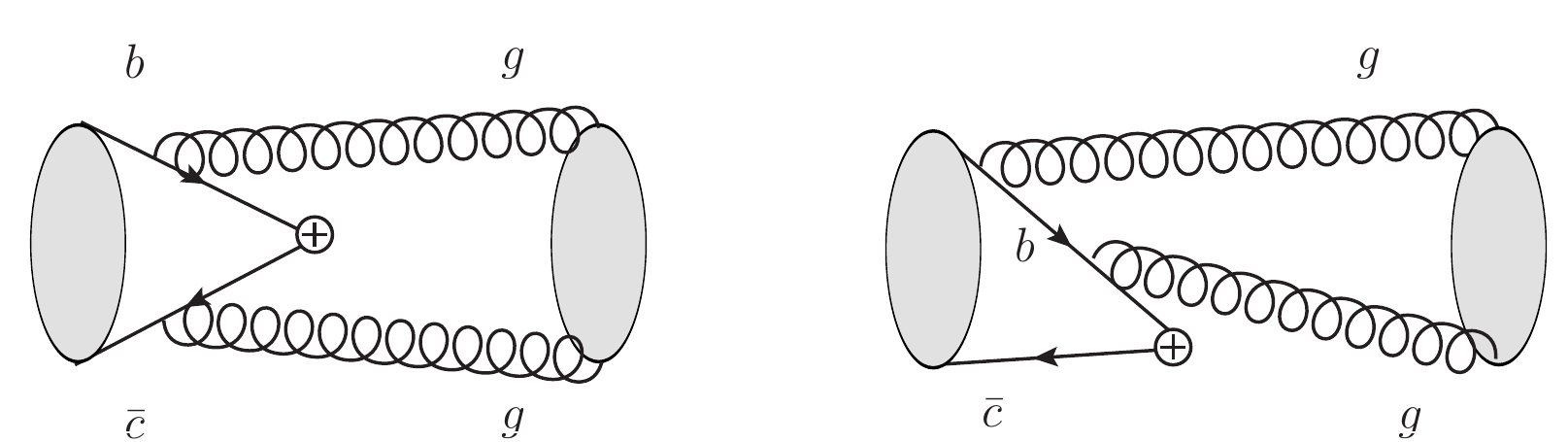}
\end{center}
    \vskip -0.7cm \caption{Typical Feynman diagrams for the form factors of $B_{c}$ into $\eta^{(')}$ and Glueball. }\label{Fig-formfactors2}
\end{figure}

\section{Form factors of $B_{c}$ into $\eta^{(')}$ and Glueball\label{III}}

The form factors of $B_c$ into a light pseudoscalar meson $P$, i.e. $f_{+}$ and
$f_{0}$ are defined in common~\cite{Wirbel:1985ji,Qiao:2012vt}
\begin{eqnarray}
\langle P(p)\vert \bar c \gamma^{\mu}b\vert B_{c}(p_{B_c})\rangle
&=&f^P_{+}(q^{2})(k^{'\mu}-\frac{m_{B_{c}}^{2}-
m_{P}^{2}}{q^{2}}q^{\mu})\nonumber\\ &&+f^P_{0}(q^{2})
\frac{m_{B_{c}}^{2}-m_{P}^{2}}{q^{2}}q^{\mu}\,,\label{ffs}
\end{eqnarray}
where the momentum transfer is defined as $q=p_{B_c}-p$ with the $B_c$ meson momentum $p_{B_c}$ and the
light meson momentum $p$, and the momentum $k'$ is defined as $k'=p_{B_c}+p$. For the decays of $B_c$ into $\eta^{(')}$, the form factors
$f^{\eta^{(')}}_{+}(q^{2})$ and $f^{\eta^{(')}}_{0}(q^{2})$
can be defined by the exchange of $P\to \eta^{(')}$. For the decays of $B_c$ into Glueball, the form factors $f^{G}_{+}(q^{2})$ and $f^{G}_{0}(q^{2})$
can be defined by the exchange of $P\to G$.

It is convenient to write the unintegrated form factors as
\begin{eqnarray}
&&\langle P(xp,(1-x)p)\vert \bar c \gamma^{\mu}b\vert B_{c}(p_{B_c})\rangle
\nonumber\\ &=&f^P_{+}(q^{2},x)(k^{'\mu}-\frac{m_{B_{c}}^{2}-
m_{P}^{2}}{q^{2}}q^{\mu})\nonumber\\ &&+f^P_{0}(q^{2},x)
\frac{m_{B_{c}}^{2}-m_{P}^{2}}{q^{2}}q^{\mu}\,.\label{ffs2}
\end{eqnarray}
After performing the integration, we have $f^P_{+,0}(q^2)=\int_0^1  f^P_{+,0}(q^2,x)dx$.

Typical Feynman diagrams for the form factors of $B_{c}$ into a light pseudoscalar meson $P$, i.e. $\eta^{(')}$ and Glueball
are plotted in Fig.~\ref{Fig-formfactors2}. Other Feynman diagrams can be obtained by changing the gluon vertex
to anti-charm quark line.

Considering the hard scatting mechanism of two gluons transition into a light pseudoscalar meson $P$, i.e. $\eta^{(')}$ and Glueball,
 the leading order Feynman diagrams
for $B_{c}$ into  a light pseudoscalar meson $P$  by the charged vector current can be written as
\begin{eqnarray}
  {\cal M}^\mu&=& \langle P(p)\vert \bar c \gamma^{\mu}b\vert B_{c}(p_{B_c})\rangle
  \nonumber\\&=&\langle0|\chi_c^{\dagger}\psi_b|B_c\rangle
        \mathrm{Tr}[{\cal A}^{\mu}(0)\Pi_{S=0}(k=0)] ,
\end{eqnarray}
where
$\psi_b$ and $\chi_c$ represent the Pauli spinor field that annihilates
a bottom quark and creates a charm antiquark, respectively.
\begin{widetext}
\begin{eqnarray}
  {\cal A}^{\mu}(q) &=& \frac{4\pi \alpha_s C_A C_F}{ (m_{B_c} N_c)^{1/2}}\sum_{i=q,s;g}\int \frac{d^4 \ell}{(2\pi)^4}
  \varepsilon^{\alpha\beta \rho \sigma} \ell_{\rho} (p_{P}-\ell)_{\sigma} F^{(i)}_{P g^* g^*}
   (\ell^2, (p_{P}-\ell)^2)\frac{1} {\ell^2\left(p_{P}-\ell\right)^2 } \nonumber  \\
                    && \times\Big[ \frac{\gamma^\beta
  (m_c+p\!\!\!\slash_{P}-\ell\!\!\!\slash-p\!\!\!\slash_2)\gamma^\alpha(m_c+p\!\!\!\slash_{P}-p\!\!\!\slash_2)\gamma^\mu}
  {\left(\left(p_2-p_{P}\right)^2-m_c^2\right)\left(\left(p_2+\ell-p_{P}\right)^2-m_c^2\right)}+\frac{\gamma^\mu
  (m_b+p\!\!\!\slash_1-p\!\!\!\slash_{P})\gamma^\beta(m_b+p\!\!\!\slash_1-\ell\!\!\!\slash)\gamma^\alpha}
  {\left(\left(p_1-p_{P}\right)^2-m_b^2\right)\left(\left(p_1-\ell\right)^2-m_b^2\right)} \nonumber  \\
                    && +\frac{\gamma^\beta
  (m_c+p\!\!\!\slash_{P}-p\!\!\!\slash_2-\ell\!\!\!\slash)\gamma^\mu(m_b+p\!\!\!\slash_1-\ell\!\!\!\slash)\gamma^\alpha}
  {\left(\left(p_2+\ell-p_{P}\right)^2-m_c^2\right)\left(\left(p_1-\ell\right)^2-m_b^2\right)}\Big],
  \label{Amu0}
\end{eqnarray}
\end{widetext}
where $p_1$ being the bottom quark momentum, $p_2$ being the anti-charm quark momentum, and $p_{P}$ being the momentum of the   light pseudoscalar meson $P$, i.e. $\eta$, $\eta'$and Glueball.

The Mathematica software is employed with the
help of the packages FeynCalc\cite{Mertig:1990an}, FeynArts\cite{Hahn:1998yk},
and LoopTools\cite{Hahn:2000jm} in the calculation of the form factors. In order to obtain the values of form factors at the maximum recoil point, we will adopt the parameter values as follows:
$m_{B_c}=6.276$GeV,
$m_{\eta}=547.85$MeV, $m_{\eta '}=957.78$MeV~\cite{PDG2018}. The heavy quark masses are adopted as $m_c=(1.5\pm0.1)$GeV and
 $m_b=(4.8\pm 0.1)$GeV~\cite{Zhu:2017lwi,Zhu:2017lqu}.
 The Gegenbauer momenta
are adopted as $a_2(1GeV)=a_2^{q,s}(1GeV)=0.44\pm0.22$~\cite{Xiao:2014uza} and $a_2^{g}(1GeV)=0.1$~\cite{Qiao:2014pfa}. Their values at other scale can be obtained by the scale evolution
equations in Eqs.~(\ref{gegen011}) and (\ref{gegen02}-\ref{gegen03}).
The running strong coupling constant is adopted around the $\eta'$ mass and one has $\alpha_s(1GeV)=0.42$.
If one chooses the scale at the charm quark mass with $m_c=(1.5\pm0.1)$GeV, one has $\alpha_s(m_c)=0.32-0.34$ and in this case the values of the form factors will be reduced by (30-40)\%.

\begin{table}[thb]
\caption{\label{tab:ffs} Form factors of the  $B_{c}$ into $\eta$ in the maximum  recoil point with $q^2=0$.
Here and in the following tables, the uncertainty is from the choice of the bottom and charm quark masses. Note that $f_{+}(0)=f_{0}(0)$. }
\begin{center}
\begin{tabular}{cccccc}
\hline\hline
Contributions &$ 10^{-3}f^\eta_{0}(q^{2}=0)$ \\
\hline
$q\bar{q} $ with LO Gegenbauer & $1.23^{+0.04}_{-0.05}$  \\
$q\bar{q} $ with NLO Gegenbauer& $1.38^{+0.00}_{-0.02}$ \\
\hline\hline
\end{tabular}
\end{center}
\end{table}

\begin{figure}[th]
\begin{center}
\includegraphics[width=0.42\textwidth]{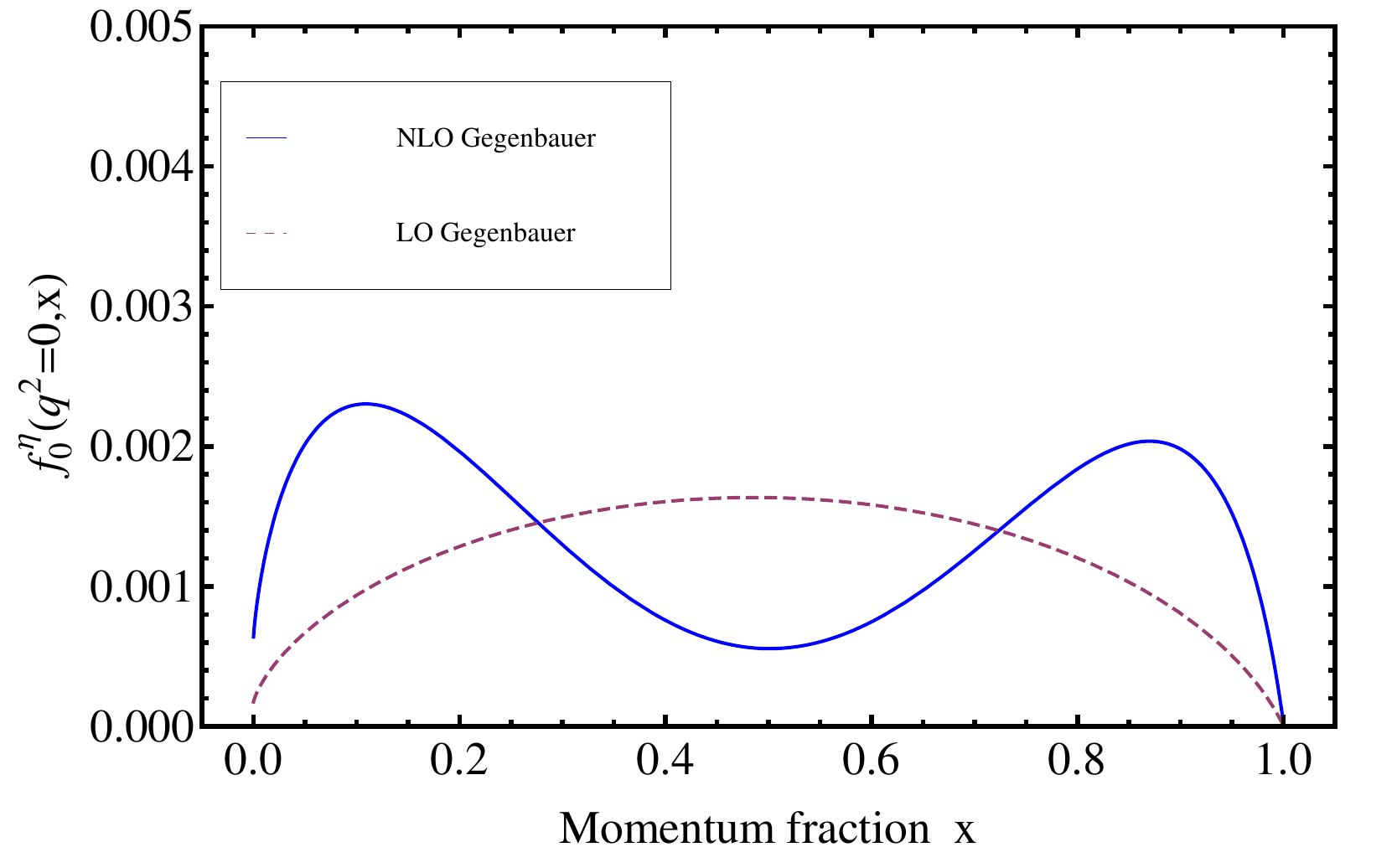}
\end{center}
    \vskip -0.7cm \caption{The unintegrated form factor of $B_{c}$ into $\eta$  dependent on the meson momentum fraction. Here
    we input the mixing angle
 $\phi=39.7^0$. Note that $f^\eta_{+}(q^{2}=0,x)\equiv f^\eta_{0}(q^{2}=0,x) $.  }\label{Fig:f0-x}
\end{figure}

\begin{figure}[th]
\begin{center}
\includegraphics[width=0.42\textwidth]{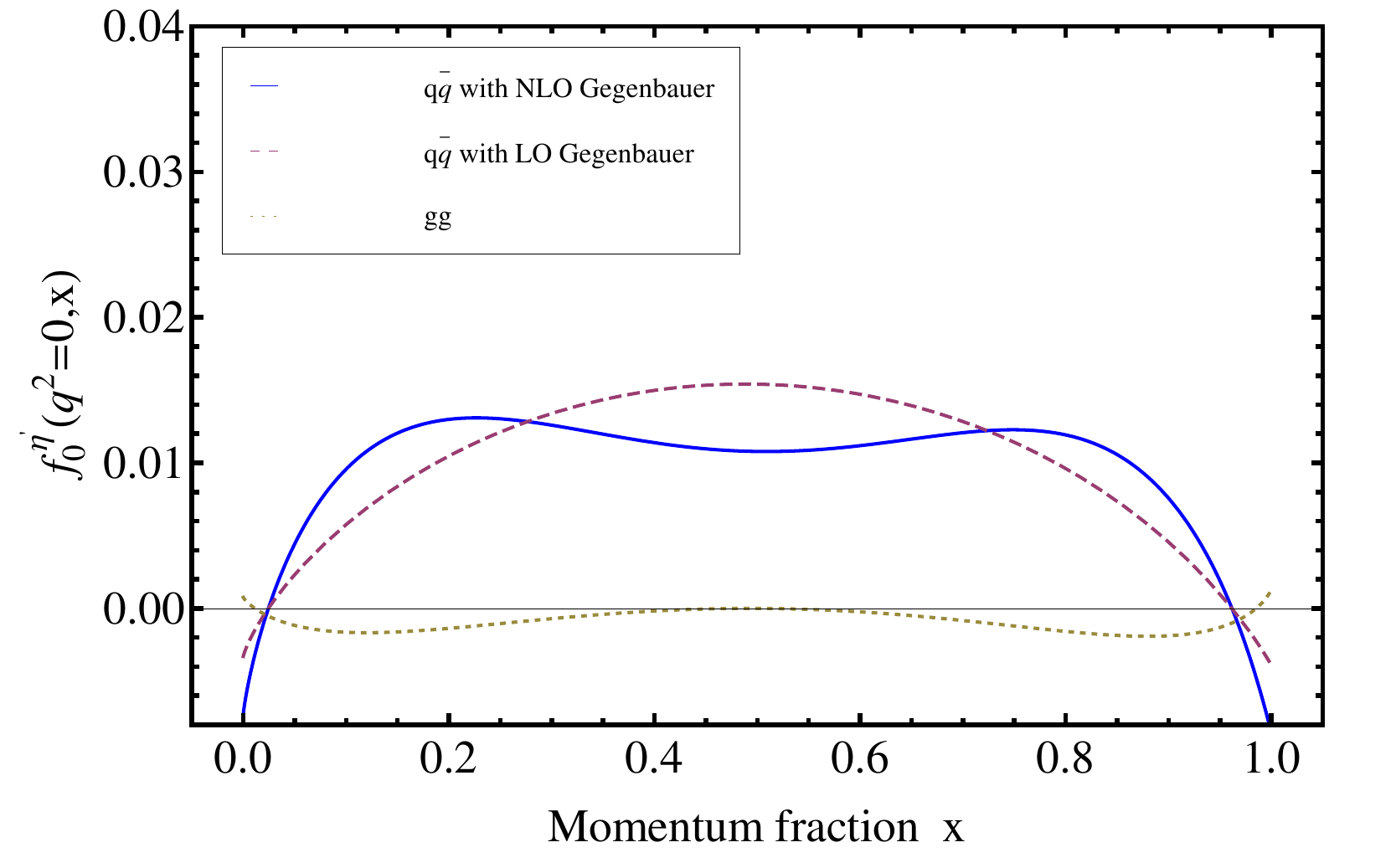}
\end{center}
    \vskip -0.7cm \caption{The unintegrated form factor of $B_{c}$ into $\eta'$  dependent on the meson momentum fraction.  Here and in the following,
    we input the mixing angles
 $\phi=39.7^0$ and $\sin^2\phi_G=0.14$~\cite{Ambrosino:2006gk}.
    Note that $f^{\eta'}_{+}(q^{2}=0,x)\equiv f^{\eta'}_{0}(q^{2}=0,x) $.  }\label{Fig:f0p-x}
\end{figure}

\begin{figure}[th]
\begin{center}
\includegraphics[width=0.42\textwidth]{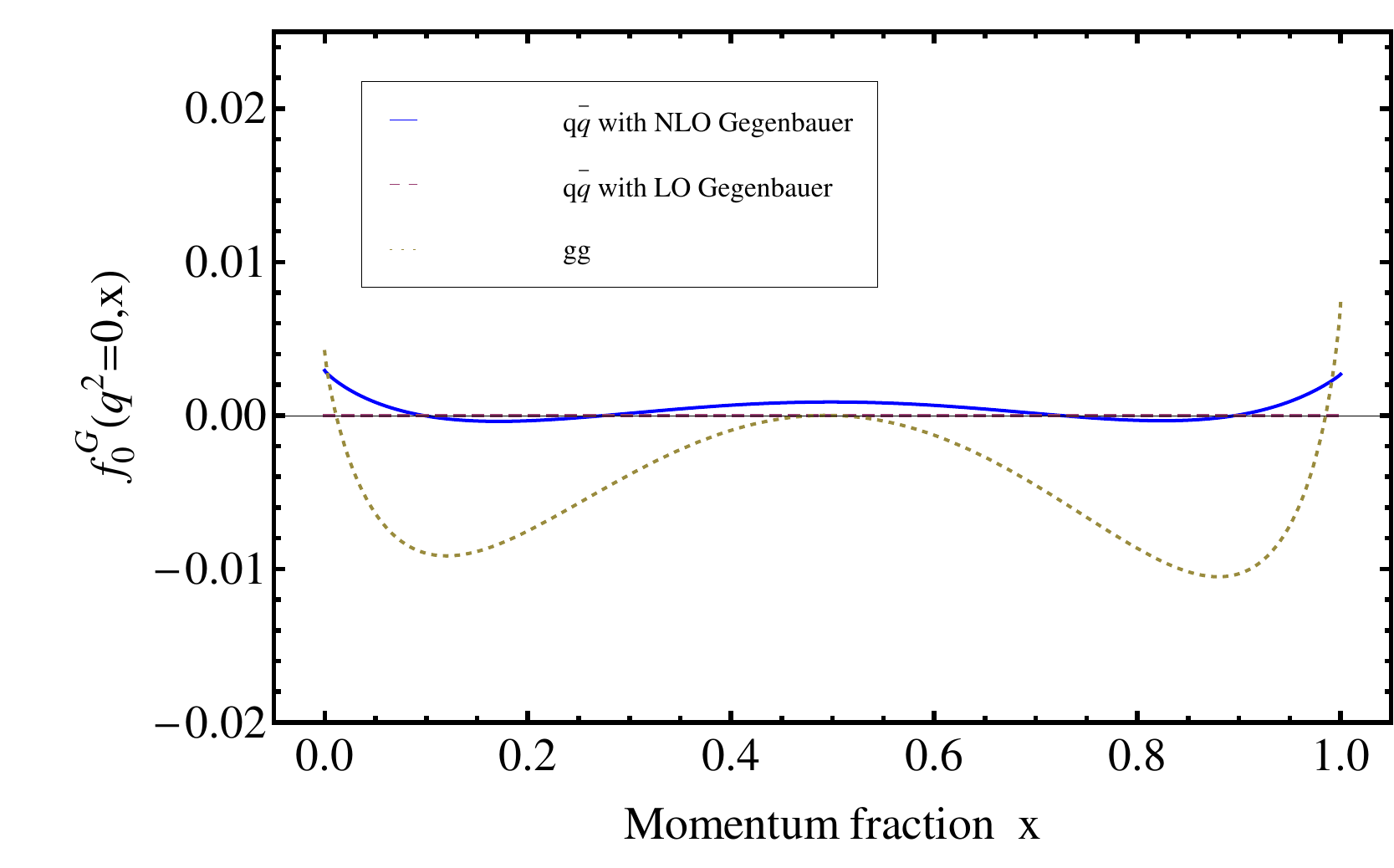}
\end{center}
    \vskip -0.7cm \caption{The unintegrated form factor of $B_{c}$ into $0^{-+}$ Glueball dependent on the meson momentum fraction. Note that $f^{G}_{+}(q^{2}=0,x)\equiv f^{G}_{0}(q^{2}=0,x) $.  }\label{Fig:f0G-x}
\end{figure}

In the maximum momentum recoil point, the form factors appearing in the expression (\ref{ffs}) can be
perturbatively calculated reliably. We give the form factors in the maximum momentum recoil point in Tabs.~\ref{tab:ffs} and~\ref{tab:ffs2},
where we input the mixing angles
 $\phi=39.7^0$ and $\sin^2\phi_G=0.14$~\cite{Ambrosino:2006gk}. When inputting other values of the mixing angles such as
 $\phi=39.7^0\pm0.7^0$ and $\sin^2\phi_G=0.14\pm0.04$ in Ref.~\cite{Ambrosino:2006gk} and $\phi=41.4^0\pm1.3^0$
 and $\sin^2\phi_G=0.04\pm0.09$ in Ref.~\cite{Escribano:2007cd}, one can easily get the corresponding values of the form factors by the
 definition of the  parameters $ C^{(q,s)}_{\eta }$, $ C^{(q,s)}_{\eta '}$, $ C^{(q,s)}_{\eta '}$, $C^{(q,s)}_{G}$ and $C^{(g)}_{G}$.
\begin{table}[thb]
\caption{\label{tab:ffs2} Form factors of the  $B_{c}$ into $\eta'$  and $0^{-+}$ Glueball
in the maximum  recoil point with $q^2=0$. Here and in the following tables, the uncertainty is from the
choice of the bottom and charm quark masses. Note that $f_{+}(0)=f_{0}(0)$. }
\begin{center}
\begin{tabular}{cccccc}
\hline\hline
Contributions &$ 10^{-2}f^{\eta'}_{0}(q^{2}=0)$ &  $ 10^{-2}f^{G}_{0}(q^{2}=0)$ \\
\hline
$q\bar{q} $ with NLO Gegenbauer& $0.97^{+0.10}_{-0.09}$ &$0.04^{+0.04}_{-0.02}$ \\
gg& $-0.08^{+0.00}_{-0.01}$ &$-0.48^{+0.09}_{-0.03}$ \\
Total & $0.89^{+0.11}_{-0.10}$ &$-0.44^{+0.13}_{-0.05}$ \\
\hline\hline
\end{tabular}
\end{center}
\end{table}

The unintegrated form factors dependent on the meson momentum fraction are sensitive to the shapes of the Gegenbauer series of the light meson.
We give the unintegrated form factors dependent on the meson momentum fraction
in Figs.~\ref{Fig:f0-x}, \ref{Fig:f0p-x}  and \ref{Fig:f0G-x}.
For the quarks contents contributions, the momentum fraction dependent shapes of  form factors  with
 leading order (LO) Gegenbauer momentum have only one peak, while that of  form factors
  with  next-to-leading (NLO)  Gegenbauer momentum will have two peaks. From Fig.~\ref{Fig:f0p-x}, ones see that the gluonium content will contribute
  the form factors of $B_c$ into $\eta'$.
From Fig.~\ref{Fig:f0G-x}, the quark contents will contribute
  the form factors of $B_c$ into Glueball.

\begin{figure}[th]
\begin{center}
\includegraphics[width=0.42\textwidth]{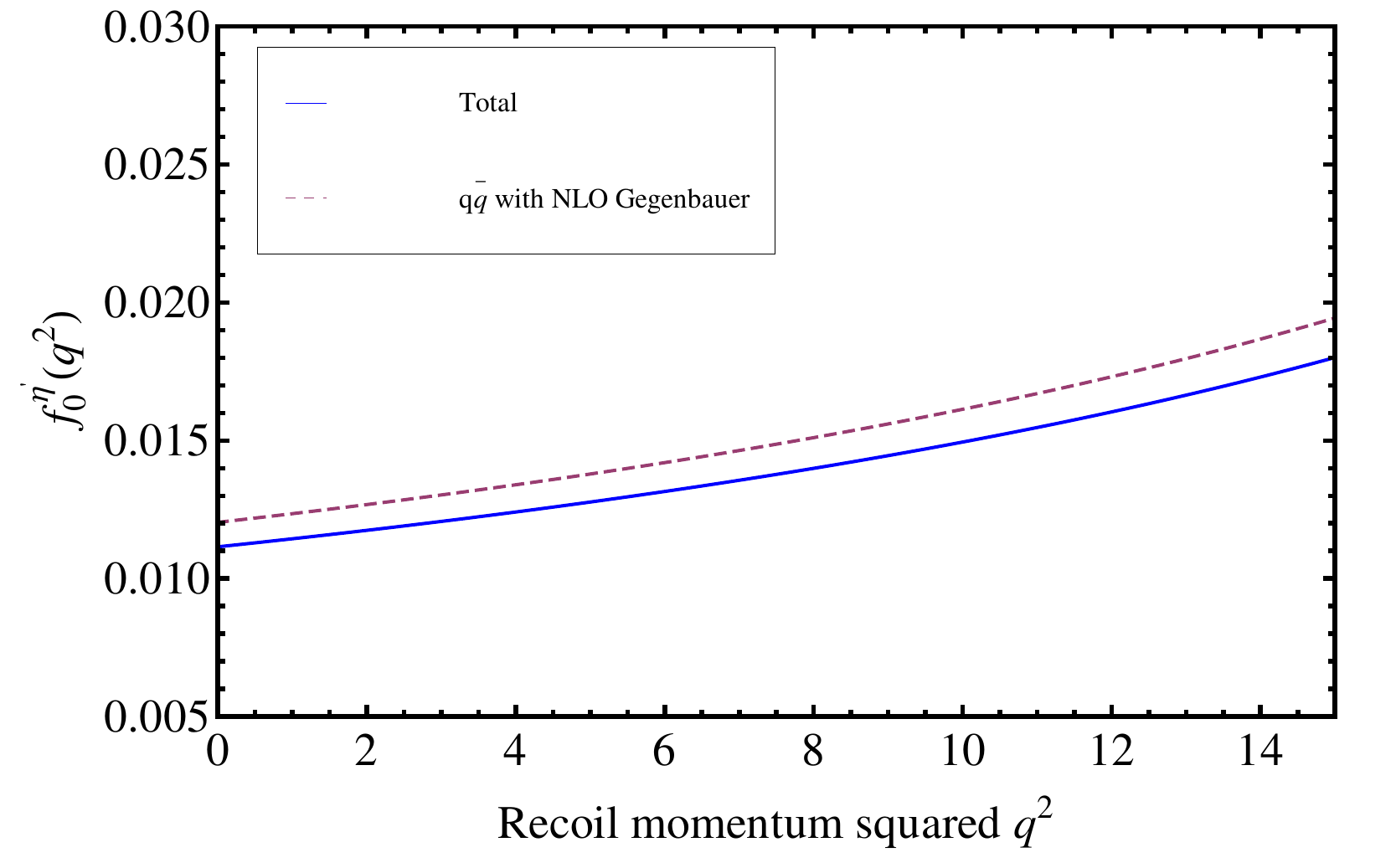}
\end{center}
    \vskip -0.7cm \caption{The form factor of $B_{c}$ into $\eta$  dependent on the recoil momentum squared.   }\label{Fig:f0-q2}
\end{figure}
\begin{figure}[th]
\begin{center}
\includegraphics[width=0.42\textwidth]{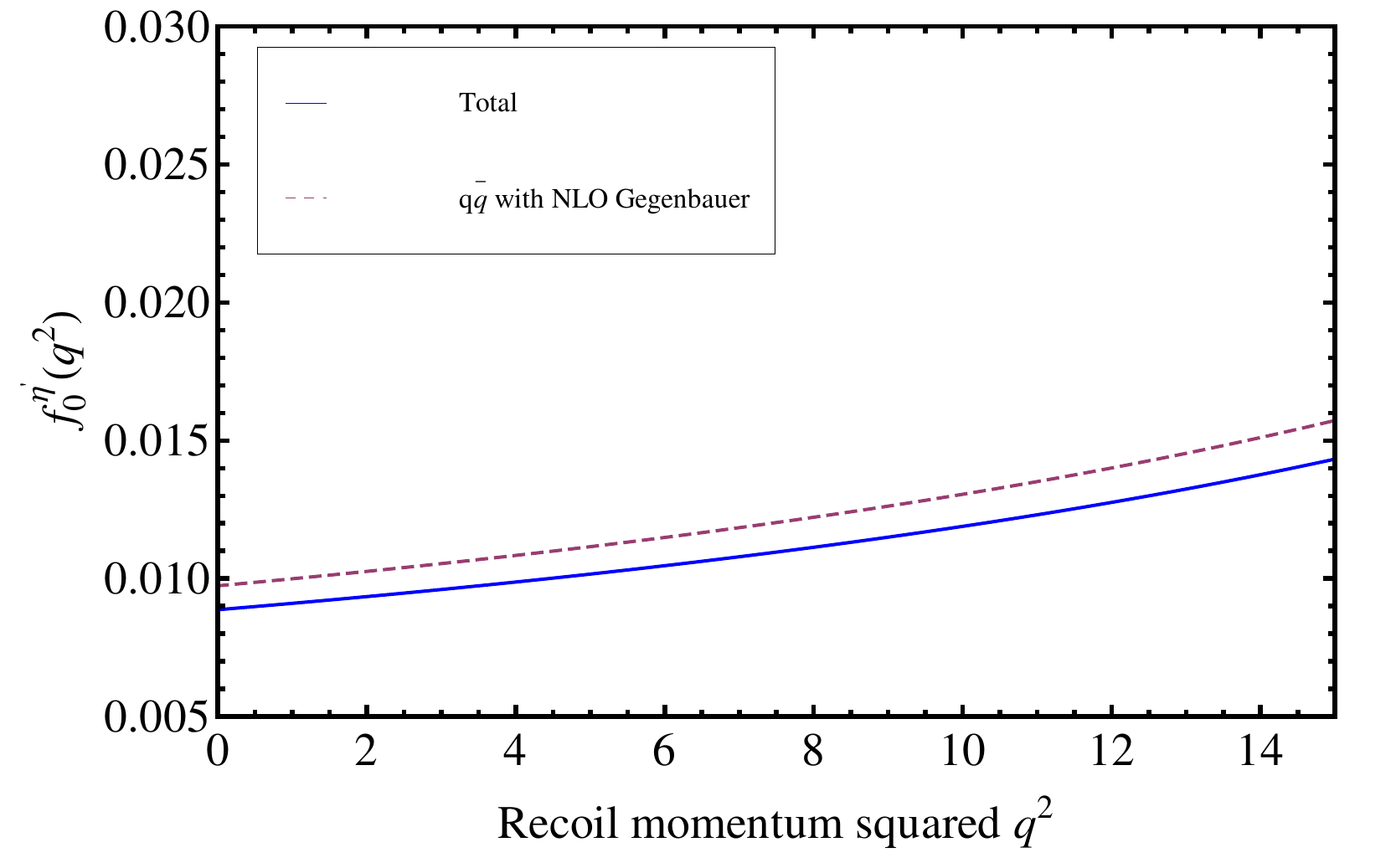}
\end{center}
    \vskip -0.7cm \caption{The form factor of $B_{c}$ into $\eta'$ dependent on the recoil momentum squared.  }\label{Fig:f0p-q2}
\end{figure}

\begin{figure}[th]
\begin{center}
\includegraphics[width=0.42\textwidth]{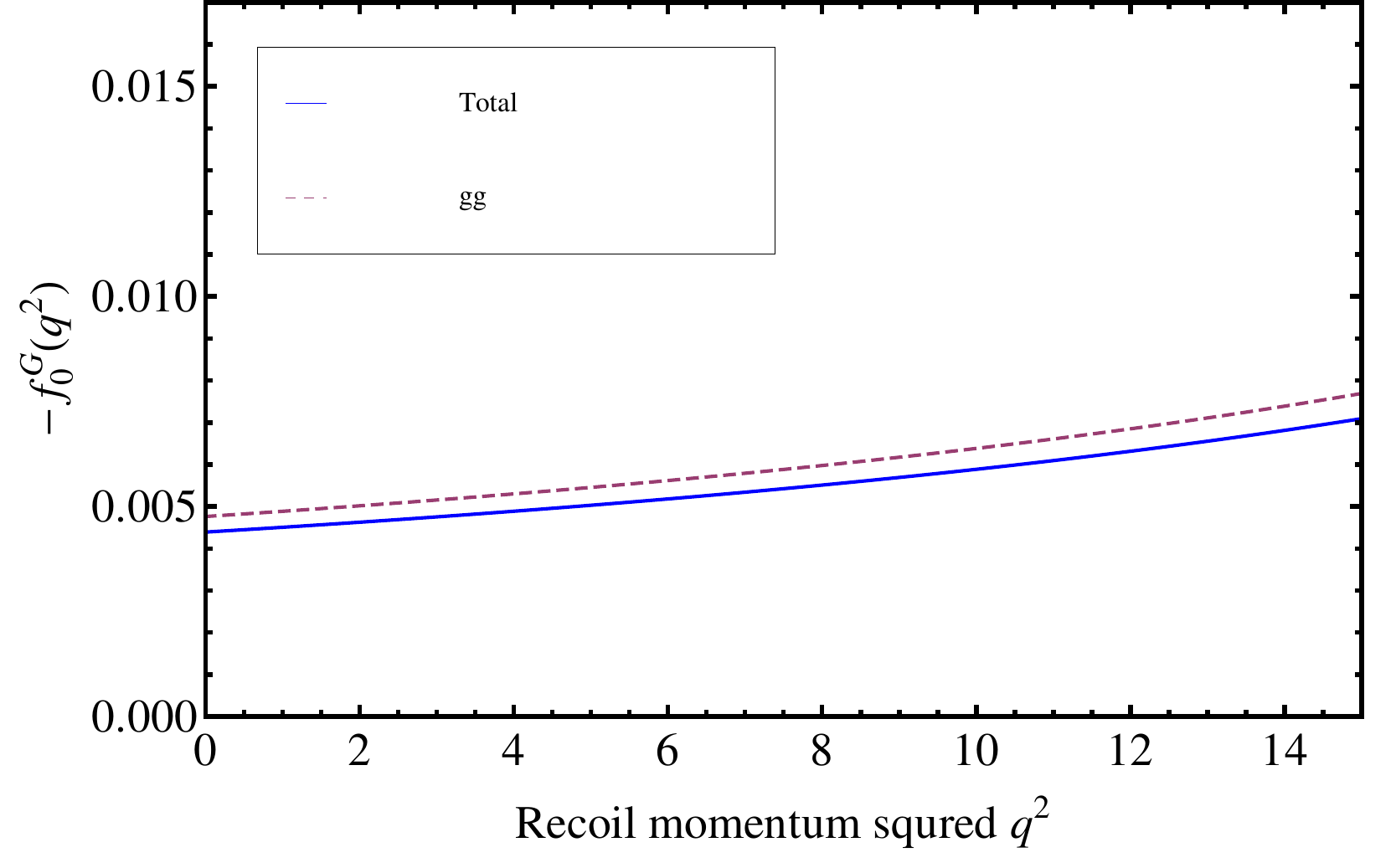}
\end{center}
    \vskip -0.7cm \caption{The form factor of $B_{c}$ into $0^{-+}$ Glueball   dependent on the recoil momentum squared. Considering that
    the value of the form factors is negative, we denoted ``$-f^G_0(q^2)$'' in the longitudinal coordinates.  }\label{Fig:f0G-q2}
\end{figure}

At the minimum momentum recoil point, the perturbative calculations for the form factors become invalid. In these region, one has to refer to
 Lattice QCD simulations or some certain models. In order to extrapolate the form factors to the minimum momentum recoil region,
the pole  model are generally adopted in many literatures~\cite{Verma:2011yw,Duplancic:2015zna}. Thus the $q^2$ distribution of the form factors can be
parametrized as
\begin{equation}\label{pole mass}
    f^P_{0,+}(q^2)=\frac{f^P_{0,+}(0)}{(1-\frac{q^2}{m^2_{B_c}})(1-a\frac{q^2}{m^2_{B_c}}+b
   \frac{ q^4}{m^4_{B_c}})}\; ,
\end{equation}
where  the $a$ and $b$ are model independent  parameters and can be fitted when the data are available. Here we let $a=b=0$ for simplication.
We plotted the form factors dependent on the  momentum transfer squared in Fig.~\ref{Fig:f0-q2}, \ref{Fig:f0p-q2} and \ref{Fig:f0G-q2}.

\section{Semileptonic decays  of $B_{c}$ into $\eta^{(')}$ and Glueball\label{IV}}

\begin{table}[th]
\begin{center}
\caption{The semileptonic decay widthes and the branching ratios of $B_{c}$ into $\eta$. Here and in the following $q\bar{q}$ denote all the quark contents, and the life time $\tau_{B_c}=0.50ps$.}
\label{tab-eta}
\begin{tabular}{c|c|c|c}
 \hline\hline
   \multicolumn{2}{c|}{$B_c \to \eta +\ell+\bar{\nu}_\ell$}& $\Gamma_{\eta}(\times 10^{-19} GeV)$ & $Br_{\eta}(\times 10^{-7})$  \\
 \hline
 \multirow{2}{*}{$\ell=e,\mu$}& $q \bar{q}(LO)$
    & $2.44^{+0.15}_{-0.18}$ & $1.85^{+0.12}_{-0.13}$  \\
\cline{2-4}
     & $q \bar{q}(NLO)$
    & $3.07^{+0.00}_{-0.07}$ & $2.33^{+0.00}_{-0.05}$  \\\hline
 \multirow{2}{*}{$\ell=\tau$}& $q \bar{q}(LO)$
     & $2.37^{+0.15}_{-0.17}$ & $1.80^{+0.11}_{-0.13}$ \\

  \cline{2-4}
   & $q \bar{q}(NLO)$
     & $2.99^{+0.00}_{-0.08}$ & $2.27^{+0.00}_{-0.05}$\\
    \hline\hline
\end{tabular}
\end{center}
\end{table}

\begin{table}[th]
\begin{center}
\caption{The semileptonic decay widthes and the branching ratios of $B_{c}$ into $\eta'$.}
\label{tab-etap}
\begin{tabular}{c|c|c|c}
 \hline\hline
   \multicolumn{2}{c|}{$B_c \to \eta' +\ell+\bar{\nu}_\ell$}& $\Gamma_{\eta^{'}}(\times 10^{-17} GeV)$ & $Br_{\eta^{'}}(\times 10^{-5})$  \\
 \hline
 \multirow{3}{*}{$\ell=e,\mu$}& $q \bar{q}(NLO)$
    &$1.24^{+0.28}_{-0.21}$ & $0.94^{+0.21}_{-0.18}$  \\
\cline{2-4}
     & $total$
     & $1.02^{+0.24}_{-0.16}$&  $0.78^{+0.18}_{-0.12}$  \\
   \hline
    \multirow{3}{*}{$\ell=\tau$}& $q \bar{q}(NLO)$
     & $1.04^{+0.23}_{-0.18}$&  $0.79^{+0.23}_{-0.13}$\\
   \cline{2-4}
     & $total$
     & $0.86^{+0.19}_{-0.14}$ &  $0.65^{+0.14}_{-0.11}$  \\
    \hline\hline
\end{tabular}
\end{center}
\end{table}

\begin{table}[th]
\begin{center}
\caption{The semileptonic decay widthes and the branching ratios of $B_{c}$ into $0^{-+}$ Glueball.}
\label{tab-etaG}
\begin{tabular}{c|c|c|c}
 \hline\hline
   \multicolumn{2}{c|}{$B_c \to G(0^{-+}) +\ell+\bar{\nu}_\ell$}& $\Gamma_{G}(\times 10^{-18} GeV)$ & $Br_{G}(\times 10^{-6})$  \\
 \hline
 \multirow{3}{*}{$\ell=e,\mu$}& $gg$
    &$2.21^{+0.15}_{-0.78}$ & $1.68^{+0.14}_{-0.60}$  \\
\cline{2-4}
     & $total$
     & $2.03^{+0.00}_{-0.69}$&  $1.54^{+0.00}_{-0.52}$  \\
   \hline
    \multirow{3}{*}{$\ell=\tau$}& $gg$
     & $1.59^{+0.14}_{-0.57}$  & $1.21^{+0.10}_{-0.43}$\\
   \cline{2-4}
     & $total$
     & $1.46^{+0.00}_{-0.49}$ &  $1.11^{+0.00}_{-0.38}$  \\
    \hline\hline
\end{tabular}
\end{center}
\end{table}

In this section, we will employ the form factors into the semileptonic decays of $B_{c}$ into $\eta^{(')}$ and Glueball.  We remain the leptons
masses, and the semileptonic partial decay width of $B_{c}$ into $\eta$ can be written as

\begin{eqnarray}
  \label{eq:dg}
\frac{{\rm d}\Gamma}{{\rm d}q^2}&=&\frac{G_F^2
\lambda(m_{B_c}^2,m_\eta^2,q^2)^{1/2}|V_{cb}|^2}{384\pi^3 m_{B_c}^3}\left(\frac{q^2-m_\ell^2}{q^2}\right)^2\frac{1}{q^2}
\nonumber\\
&&\times[(m_\ell^2+2q^2)\lambda(m_{B_c}^2,m_\eta^2,q^2)(f^\eta_+(q^2))^2\nonumber
\\&&+3m_\ell^2(m_{B_c}^2-m_\eta^2)^2(f^\eta_0(q^2))^2]\,,
\end{eqnarray}
where $\lambda(m_{B_c}^2,m_\eta^2,q^2)=(m_{B_c}^2+m_{\eta}^2-q^2)^2-4m_{B_c}^2
m_{\eta}^2$. And the similar formulae can be obtained for the semileptonic decay width of $B_{c}$ into $\eta'$ and Glueball with the replacement of $\eta\to\eta'(G)$.

The semileptonic decay widthes and the branching ratios can be obtained after integrating the momentum recoil squared $q^2$.
In Tab.~\ref{tab-eta}, we give  the results for the $B_c \to \eta +\ell+\bar{\nu}_\ell$ with $\ell=e,~\mu,~\tau$. The masses of the leptons are:
$m_{e}=0.50$MeV, $m_{\mu}=105.6$MeV and $m_{e}=1777$MeV~\cite{PDG2018}. The form factors
 with  both LO and NLO Gegenbauer series are considered in the semileptonic decays.  From the table, their decay widthes are around
$ 10^{-19}GeV$, while the branching ratios are around $ 10^{-7}$.
For $\ell=e,~\mu$, their decay widthes are nearly identical since their masses
can be discarded in the $B_c$ meson decays to $\eta$. Besides, the  LO and NLO Gegenbauer series  have less
influence in the semileptonic decay width of $B_{c}$ into $\eta$.
In Tab.~\ref{tab-etap}, we give  the results for the $B_c \to \eta' +\ell+\bar{\nu}_\ell$ with $\ell=e,~\mu,~\tau$. The form factors
 from the quark content with  NLO Gegenbauer series  and from the gluonium contribution are considered
 in the semileptonic decays.   From the table, their decay widthes are around
$ 10^{-17}GeV$, while the branching ratios are around $ 10^{-5}$.   We give  the results for the $B_c \to G(0^{-+}) +\ell+\bar{\nu}_\ell$
in Tab.~\ref{tab-etaG}, where the decay widthes are around
$ (10^{-18}-10^{-17})GeV$, while the branching ratios are around $ (10^{-6},10^{-5})$.

In Ref.~\cite{Gang:2006nc}, the semileptonic branching ratios of $B_{c}$ into $\eta^{(')}$ have been already predicted in perturbative QCD,
where the $Br(B_c \to \eta +\ell+\bar{\nu}_\ell)=3.98\times 10^{-6}$ and  $Br(B_c \to \eta' +\ell+\bar{\nu}_\ell)=5.24\times 10^{-5}$ with $\ell=e,\mu$ and $m_u=2$MeV, $m_d=4$MeV and $m_s=80$MeV.
Compared with these predictions in Perturbative QCD, our results are smaller due to
the choice of the decay constant of $\eta^{(')}$ and the two gluon scattering mechanism is employed.
 Currently, there is no report on semileptonic decays of $B_{c}$ into $\eta^{(')}$.
 However, the hunting for the signals of $B_{c}$ into $\eta^{(')}$ is accessible in
 future LHCb experiments when considering the large cross section of $B_c$ meson.

In the end it is very interesting to find out whether the formulae in above can guide
the studies of the processes $D_s\to\eta +\ell+\bar{\nu}_\ell$. The BESIII Collaboration have measured
these channels and given  $Br(D_s\to\eta +\ell+\bar{\nu}_\ell)=(2.42\pm0.46\pm0.11)\%$ and  $Br(D_s\to\eta' +\ell+\bar{\nu}_\ell)=(1.06\pm0.54\pm0.07)\%$~\cite{Ablikim:2017omq}.  For $D_s\to\eta^{(')} +\ell+\bar{\nu}_\ell$, the $c\to s$ transition with another spectator strange quark will be present in the $D_s\to\eta^{(')}$ form factors. Considering the transferred momentum is small in $D_s\to\eta' +\ell+\bar{\nu}_\ell$, the perturbative calculation may be invalid, so we only consider the channel $D_s\to\eta +\ell+\bar{\nu}_\ell$. As the tentatively analysis, it is interesting to find out how large the mechanism of  two gluon transitions contributes to processes $D_s\to\eta +\ell+\bar{\nu}_\ell$. Employing the above formulae, and taking the replacement of $b\to c$, $c\to s$ and $B_c \to D_s$, we may  tentatively give the order of magnitude of their decay widthes since the $D_s$ meson is not a really nonrelativistic bound state.  We found that the mechanism of two gluon transitions  gives $Br(D_s\to\eta +\ell+\bar{\nu}_\ell)\sim 10^{-4}$  and  only contributes to   0.5\% in the channel  $D_s\to\eta +\ell+\bar{\nu}_\ell$. The $c\to s$ transition  thus dominates the form factor of  $D_s\to\eta$. To extrapolate the form factors of  $D_s\to\eta$ to the minimum momentum recoil region,
the pole  model is still useful~\cite{Duplancic:2015zna,Cheng:2017pcq}. Combing the experimental data, the $c\to s$ transition leads to the $
 f^{D_s\eta}_{0,+}(q^2=0) =0.50\pm0.05$.

\section{Conclusion}

We investigated the form factors of  $B_{c}$ into the $\eta^{(')}$ and pseudoscalar Glueball and employed
the form factors into their semileptonic decays.
Unlike the decay of $D_{s}$ into $\eta^{(')}$ where the $c\to s$ transition is dominant, the two-gluon scattering mechanism dominated the contribution for the form factors
of  $B_{c}$ into $\eta^{(')}$. We considered the  $\eta -\eta '-$Glueball mixing effects, and studied their influences in the form factors.
 At the maximum momentum recoil point, the form factors of  $B_{c}$ into the light pseudoscalar mesons are factored as
  the LDMEs along with the corresponding  short-distance perturbatively calculable coefficients.
  The  results of form factors  in the maximum momentum  recoil point were obtained.
   Using the pole model, the  form factors of  $B_{c}$ into  the $\eta^{(')}$ and pseudoscalar Glueball are extrapolated into the minimum momentum recoil region.

The corresponding semileptonic decay widthes and the branching ratios were calculated.
The results are: the branching ratio of $B_c \to \eta +\ell+\bar{\nu}_\ell$ is around $ 10^{-7}$;
 the  branching ratio of $B_c \to \eta' +\ell+\bar{\nu}_\ell$ is around $ 10^{-5}$; the  branching ratio of  $B_c \to G(0^{-+}) +\ell+\bar{\nu}_\ell$ is around $ (10^{-6},10^{-5})$.
Future LHCb experimental shall test these predictions, which is helpful to understand the
  $\eta-\eta'-$Glueball  mixing and the decay properties of $B_c$ meson.

%%%%%%%%%%%%%%%%
%%%%%%%%%%%%%%%%
\section*{Acknowledgments}
This work was supported in part by the National Natural Science Foundation
of China under Grant No. 11705092, 11775117 and 11235005, and by Natural Science Foundation of
Jiangsu under Grant No.~BK20171471.

\end{document}